\documentclass[apj,numberedappendix]{emulateapj}
\usepackage{apjfonts}

\newcommand{\be}{\begin{eqnarray}}
\newcommand{\ee}{\end{eqnarray}}

\newcommand{\lp}{\left(}
\newcommand{\rp}{\right)}
\newcommand{\lb}{\left[}
\newcommand{\rb}{\right]}

\begin{document}

\slugcomment{Submitted for publication in The Astrophysical Journal}

\shorttitle{Radioactively-Powered Rising Lightcurves of SNe Ia}
\shortauthors{Piro, A. L.}

\normalsize


\title{Radioactively-Powered Rising Lightcurves of Type Ia Supernovae}

\author{Anthony L. Piro}

\affil{Theoretical Astrophysics, California Institute of Technology, 1200 E California Blvd., M/C 350-17, Pasadena, CA 91125; piro@caltech.edu}


\begin{abstract}
The rising luminosity of the recent, nearby supernova 2011fe shows a quadratic dependence with time during the first $\approx0.5-4\ {\rm days}$. In addition, the composite lightcurves formed from stacking together many Type Ia supernovae (SNe Ia) show a similar power-law index of $1.8\pm0.2$ with time. I explore what range of power-law rises are possible due to the presence of radioactive material near the surface of the exploding white dwarf (WD). I summarize what constraints such a model places on the structure of the progenitor and the distribution and velocity of ejecta. My main conclusion is that the rise of SN 2011fe requires a mass fraction $X_{56}\approx 3\times10^{-2}$ of $^{56}$Ni (or some other heating source like $^{48}$Cr) distributed between a depth of $\approx4\times10^{-3}-0.1M_\odot$ below the WD's surface. Radioactive elements this shallow are not found in simulations of a single C/O detonation. Scenarios that may produce this material include helium-shell burning during a double-detonation ignition, a gravitationally confined detonation, and a subset of deflagration to detonation transition models. In general, the power-law rise can differ from quadratic depending on the details of the event, so comparisons of this work with observed {\it bolometric} rises of SNe Ia would place strong constraints on the distribution of shallow radioactive material, providing important clues for identifying the elusive progenitors of SNe Ia.
\end{abstract}

\keywords{hydrodynamics ---
	shock waves ---
	supernovae: general ---
	white dwarfs}


\section{Introduction}

The use of Type Ia supernovae (SNe Ia) as cosmological distance indicators \citep{rie98,per99} has brought attention to the theoretical uncertainties that remain about these events. It is generally accepted that these SNe result from unstable thermonuclear ignition of degenerate matter \citep{hf60} in a C/O white dwarf (WD), but frustratingly the specific progenitor systems have not yet been identified. The main three candidates are (1) stable accretion from a non-degenerate binary companion until the Chandrasekhar limit is reached \citep{wi73}, (2) the merging of two C/O WDs \citep{it84,web84}, or (3) accreting and detonating a helium shell on a C/O WD that leads to a prompt detonation of the core \citep{ww94a,la95}. A variation on the latter case is ignition triggered by a detonation in an accretion stream \citep{gul10,dan12}. In addition, it is not known whether the incineration itself proceeds as a sub-sonic deflagration \citep{nom76,nom84} or deflagration-detonation transition \citep[DDT;][]{kho91,ww94b}. Single detonations of a sub-Chandrasekhar WD have been shown to reproduce many features of SNe Ia \citep{sim10}, but it is not clear how to ignite these cores without first detonating a helium shell. Each combination of situations has different implications for the velocity and density structure of the exploding WD, as well as the distribution of ashes.
   
A potentially powerful method for constraining between these models are comparisons with the early-time behavior of SNe Ia, since this is when the surface layers of the WD are being probed by the observed emission. The recent, nearby SN 2011fe is especially useful in this regard, since it was detected just $\approx11$ hours post explosion when the luminosity was merely $\sim10^{-3}$ of that at peak \citep{nug11}. Furthermore, upper limits on the luminosity were placed $\approx4$ hours post explosion \citep{blo12}. An interesting feature of the rise is a $t^2$ dependence for the luminosity up to $\approx5\ {\rm days}$ post explosion. Furthermore, there was no sign of the cooling of shock heated surface layers \citep{pir10,rab11} nor interaction with a companion \citep{kas10}. This puts tight constraints on the progenitor radius of $\lesssim0.02R_\odot$ \citep{blo12}, demonstrating that it was very compact and consistent with a WD. Other individual SNe Ia have not been studied in the detail of SN 2011fe, but the composite lightcurves formed from stacking many supernovae show a power-law index of $1.8\pm0.2$ \citep{con06}. Although this is roughly quadratic, it could instead indicate some diversity in the rise.

A $t^2$ dependence is consistent with a model in which the effective temperature remains fixed while the radius increases with time at constant velocity \citep{rie99}. This seems unlikely to hold for a real supernova, since the effective temperature and photospheric radius can potentially change as the ejecta expands and its density drops. In the Supplementary Information of \citet{nug11}, a single-zone model is described \citep[using arguments similar to that in][]{arn82}, estimating the luminosity from expanding ejecta that is heated by $^{56}$Ni decay. For a explosion energy $E_{\rm sn}$ and total stellar mass $M_*$, one can define a characteristic velocity of $V=(2E_{\rm sn}/M_*)^{1/2}$. Using a constant opacity $\kappa$, this gives a luminosity of
\be
	L(t) \approx \frac{2\pi}{3}\frac{cVX_{56}\epsilon_{\rm 56}t^2}{\kappa},
	\label{eq:singlezone}
\ee
where $c$ is the speed of light, $X_{56}$ is the $^{56}$Ni mass fraction, and $\epsilon_{56}=3.9\times10^{10}\ {\rm erg\ g^{-1}\ s^{-1}}$ is the radioactive heating rate per unit mass.  This result demonstrates that radioactive heating can in principle also provide a $t^2$ dependence. What this single-zone model cannot answer is what is the required depth of the radioactive material. In addition, the power-law dependence on time may change depending on gradients in density, velocity, and distribution of radioactive isotopes.

In the following work I generalize this single-zone model to a one-dimensional calculation that includes the above mentioned complications. In \S \ref{sec:dynamics}, I summarize the main equations used to describe the dynamics and thermodynamics of the heated, expanding ejecta. In \S \ref{sec:energyanddiffusion}, I discuss the time-evolving energy density of the expanding layers, and show why heating from radioactive decay dominates over the shock heating that has been the focus of previous work \citep{pir10,rab11}. I also explore the depth of the thermal diffusion wave as the surface layers expand, which shows the depth in the WD that the early lightcurve is probing as a function of time. In \S \ref{sec:luminosity}, I calculate the time-dependent luminosity, and explore what range of power-law scalings are expected for the rise. In \S \ref{sec:compare}, I compare these calculations with the observations of SN 2011fe to constrain the distribution and amount of radioactive material. Finally, in \S \ref{sec:conclusions} I summarize my results and discuss what detailed modeling can do for the understanding of the outer ejecta of SNe Ia.

\section{Dynamics and Thermodynamics of the Expanding Star}
\label{sec:dynamics}

For this study I concentrate on the plane-parallel surface layers of the exploding progenitor star. This does not mean the expansion itself is plane-parallel, but merely that all the ejecta originates from roughly the same radius. This simplification is an accurate representation of the outer material on a WD. In the Appendix, I summarize the scalings for a non-plane-parallel treatment.

Variables of the pre-expanded progenitor star are denoted with the subscript $0$. The profile is assumed to be a polytrope, with
\be
	P_0 = K \rho_0^{1+1/n},
	\label{eq:polytrope}
\ee
where in the case of non-relativistic electrons $n=3/2$ and $K=9.91\times10^{12}\mu_e^{-5/3}$, and for relativistic electrons $n=3$ and $K=1.23\times10^{15}\mu_e^{-4/3}$, where $\mu_e$ is the molecular weight per electron and $K$ is in cgs units. For a constant flux, ideal gas dominated, plane-parallel atmosphere $n=3$ and $K=6.1\times10^{13}\ g_9^{-1/3}T_{\rm eff,5}^{4/3}$, where $g=10^9g_9\ {\rm cm\ s^{-2}}$ is the surface gravity and $T_{\rm eff}=10^5T_{\rm eff,5}\ {\rm K}$ is the effective temperature of the photosphere \citep{pir10}. Depending on the progenitor model, or heating from sub-sonic burning during the explosion, a higher $T_{\rm eff}$ may be a more realistic choice \citep{rab11}.

For a typical Chandrasekhar-mass progenitor, $n=3/2$ is most relevant since at $\approx0.5\ {\rm day}$ post explosion (when SN 2011fe was first observed), the diffusion wave is at a depth where the progenitor star equation of state is set by non-relativistic electrons. Nevertheless, I keep the polytropic index as a free variable since it can vary in other situations. For example, if the WD was ignited from a detonating helium shell, the layer may have a convective profile instead \citep{bil07,sb09,wk11,she11}.

The velocity of the shock has a gradient with density, which rises toward the surface according to \citep{mm99},
\be
	V_s(\rho_0) \approx 0.79 \lp \frac{E_{\rm sn}}{M_*}\rp^{1/2} \lp\frac{M_*}{\rho_0 R_*^3} \rp^\beta,
	\label{eq:vs}
\ee
where $R_*$ is the WD radius and $\beta= 0.19$ for a radiation pressure dominated shock \citep{sak60}. I keep $\beta$ as a free parameter, so that my results are general with respect to velocity gradients that are from non-shock sources. Typical values for equation (\ref{eq:vs}) are $E_{\rm sn}\approx10^{51}\ {\rm erg}$ and the mass and radius of a WD near the Chandrasekhar limit, but it may also be worth considering a lower mass WD in light of the pure detonation simulations of  sub-Chandrasekhar explosion models \citep{sim10}. In \citet{pir10}, we focused on the shock from a DDT, and wrote the velocity gradient as
\be
	V_s(\rho_0)=V'\lp \rho_0/\rho' \rp^{-\beta}.
\ee
where $V'$ and $\rho'$ are set by where the detonation fails and a shock runs away, heating the surface of the star. This roughly gives $\rho'\approx2\times10^6\ {\rm g\ cm^{-3}}$ and $V'\approx8\times10^8\ {\rm cm\ s^{-1}}$ \citep{pir10}. For the present work I quote analytic results in terms of $\rho'$ and $V'$ since it gives the cleanest solutions. But for the numerical factors, I substitute $\rho'=M_*/R_*^3$ and $V'=0.79(E_{\rm sn}/M_*)^{1/2}$, since using $E_{\rm sn}$ and $M_*$ makes it easier to compare with other theoretical work and observations.

In this framework, one can think of the surface layers as a series of shells, each labeled with an initial density $\rho_0$ and moving with a final velocity of $V(\rho_0)\approx 2V_s(\rho_0)$ \citep[this factor of 2 is  the effect of pressure gradients causing acceleration,][]{mm99}. For a plane-parallel configuration, mass conservation as the shell expands gives
\be
	\rho(\rho_0,t) = \rho_0 \lb\frac{R_*}{V(\rho_0)t} \rb^2 \lb \frac{H_0}{\Delta V(\rho_0) t} \rb,
	\label{eq:massconservation}
\ee
where $H_0=P_0/\rho_0 g = K\rho_0^{1/n}/g$ is the thickness of the layer, which I set to be the pressure scale height, and the velocity gradient of the layer is
\be\label{eq:deltav}
	\Delta V(\rho_0)
	\approx \frac {\partial V}{\partial \rho_0}
		\frac{\partial \rho_0}{\partial r_0}H_0 = \frac{\beta}{1+1/n}V(\rho_0).
\ee
This expression is found by using the equation of hydrostatic balance, $dP_0/dr_0 = -\rho_0 g$, and equation (\ref{eq:polytrope}). Thus $\Delta V(\rho_0)$ is smaller than $V(\rho_0)$ by a constant factor of $\beta/(1+1/n)\approx0.11$ (for $n=3/2$ and $\beta=0.19$).

The thermal evolution of the expanding layer is set by the first law of thermodynamics,
\be
	Tds = d\lp\frac{U}{\rho} \rp+Pd\lp\frac{1}{\rho} \rp = \frac{1}{\rho}dU+\frac{4}{3}Ud\lp\frac{1}{\rho} \rp,
\ee
where $s$ is the specific entropy, $U$ is the energy density, and the right-hand side comes from assuming a radiation dominated energy density, so that $P=U/3$. Changes in entropy come from nuclear heating and radiative losses, so this can be rewritten as
\be
	\frac{1}{\rho}\frac{\partial U}{\partial t} +\frac{4}{3}U\frac{\partial}{\partial t}\lp\frac{1}{\rho} \rp = \epsilon_{56}X_{56}\lp\frac{\rho_0}{\rho_{56}} \rp^{\chi}-\frac{\partial L}{\partial M_r},
	\label{eq:energy}
\ee
where the partial derivatives in time are evaluated at a fixed mass shell, labeled by $\rho_0$. I assume that the $^{56}$Ni can potentially vary with depth, which is modeled with a characteristic density $\rho_{56}$ and a power-law index $\chi$. Such a power-law choice is not physically motivated, but is made simply to allow the deposition to vary, while still resulting in self-similar solutions.  I ignore changes to $X_{56}$ due to decay, since I am focusing on times earlier than the $^{56}$Ni half-life of $6.077\ {\rm days}$.

It is also possible that other radioactive isotopes could be present and be powering the early-time lightcurve. In particular, $^{48}$Cr has a 21.56 hr half-life with nearly 100\% electron captures to the $1^+$ excited state of $^{48}$V, which is followed by a cascade that emits an energy of 0.42 MeV. The $^{48}$V has a 15.973 day half-life, and an effective energy deposition of 2.874 MeV. Therefore the first step deposits $\approx7.5\times10^{10}\ {\rm erg\ g^{-1}\ s^{-1}}$, and the second step deposits $\approx2.9\times10^{10}\ {\rm erg\ g^{-1}\ s^{-1}}$. These are not too different than $\epsilon_{56}$, but may introduce some qualitative changes. The radioactive isotope $^{52}$Fe may also be present, but its short half-life of 8.275 hr means that it is negligible at all but the earliest times. For the main analysis of the paper, I concentrate on $^{56}$Ni, but one should keep in mind that the factor $X_{56}\epsilon_{56}$ can stand for a more complicated mixture of radioactive isotopes.

The derivative $\partial L/\partial M_r$ is the radiative cooling rate per unit mass, where
\be
	L = -\frac{4\pi r^2c}{3\kappa\rho}\frac{\partial U}{\partial r}.
	\label{eq:radiativeluminosity}
\ee
Throughout my main analysis I assume that $\kappa$ is independent of density and temperature, such as for an electron scattering opacity. This is roughly accurate for the hot surface layers, but in the Appendix I summarize the self-similar scalings that result from a more general opacity law.

\section{Energy Density and Diffusion Depth}
\label{sec:energyanddiffusion}

Using the framework summarized in the previous section, one can in principle solve equation (\ref{eq:energy}) for $U(\rho_0,t)$. This provides the entire dynamic and thermodynamic evolution of each mass shell (labeled by $\rho_0$) as a function of time. For the present work I am focused on the early lightcurve ($\approx0.5-4\ {\rm days}$ post explosion) in the limit where heating from $^{56}$Ni decay dominates. The strategy is then to assume that the nuclear heating is much greater than radiative cooling, so that $\partial L/\partial M_r$ can be ignored in equation (\ref{eq:energy}). Once this simplification is made, the equation can be easily integrated to find
\be
	U(\rho_0,t) = \frac{1+1/n}{16\beta}\frac{R_*^2K\rho'^{1+1/n}\epsilon_{56}X_{56}}{gV'^3t^2}
	\lp\frac{\rho_0}{\rho_{56}}\rp^\chi\lp\frac{\rho_0}{\rho'}\rp^{1+1/n+3\beta},
	\nonumber
	\\
	\label{eq:energydensity}
\ee
where there is a factor of $\approx1/8$ included from taking $V\approx 2V_s$. This solution implicitly assumes that the energy density deposited by the passing shock wave, which was the main source of heating considered by \citet{pir10} and \citet{rab11}, is negligible. To show that this is a reasonable approximation, consider a shock traversing a density $\rho_0$. The energy density deposited by that shock is
\be
	U_{\rm sh,0} = \frac{6}{\gamma+1}\rho_0V_s^2.
\ee
This energy density then adiabatically decreases as the layer expands, giving
\be
	U_{\rm sh}(\rho_0,t) = U_{\rm sh,0}\lb \frac{\rho(\rho_0,t)}{\rho_{\rm sh}} \rb^\gamma,
	\label{eq:energysh}
\ee
where $\gamma=4/3$ is the adiabatic exponent for radiation-dominated material and $\rho_{\rm sh}=\rho_0(\gamma+1)/(\gamma-1)=7\rho_0$ is the compressed, shocked density. Since $U_{\rm sh}\propto t^{-4}$ (from eq. [\ref{eq:energydensity}]) while $U\propto t^{-2}$ (from eq. [(\ref{eq:energydensity}]), the shock energy density falls off much faster and is quickly negligible in comparison to the radioactive heating. I compare the two energy densities in Figure 1 at three different snap-shots in time. This shows that the radioactive heating dominates at $\approx1\ {\rm day}$ unless $X_{\rm 56}\lesssim 10^{-3}$.

\begin{figure}
\epsscale{1.2}
\plotone{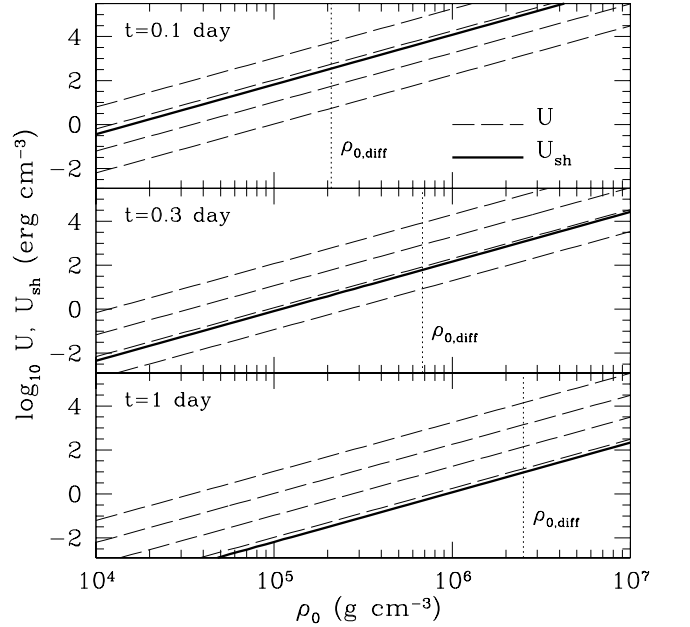}
\figcaption{Comparison of the energy density due to radioactive decay given by equation (\ref{eq:energydensity}) (dashed lines) for $X_{56}=10^{-3}, 10^{-2},10^{-1},\ {\rm and}\ 1.0$ (from bottom to top in each panel) with the energy density from shock passage given by equation (\ref{eq:energysh}) (thick, solid lines). The supernova parameters are $E_{\rm sn}=10^{51}\ {\rm erg}$, $M_*=1.4M_\odot$, and $R_*=3\times10^8\ {\rm cm}$. The vertical dotted lines show the depth of the thermal diffusion wave at each time (eq. [\ref{eq:rhodiff}]). Radioactive decay dominates the energy budget for $X_{56}\gtrsim10^{-1}$ at $2.4\ {\rm hrs}$ after shock breakout and for $X_{56}\gtrsim10^{-2}$ at $7.2\ {\rm hrs}$.}
\label{fig:energydensity}
\epsscale{1.0}
\end{figure}

Also plotted in Figure 1 is the depth of the thermal diffusion wave as the surface layers expand and cool (vertical dotted lines). The timescale for a given layer to cool is
\be
	t_{\rm diff} \approx \frac{3\kappa\rho}{c}(\Delta r)^2.
	\label{eq:tdiff}
\ee
By substituting equation (\ref{eq:massconservation}) into equation (\ref{eq:tdiff}) and setting $t_{\rm diff}=t$, I solve for the diffusion depth as a function of time,
\be
	\frac{\rho_{0,\rm diff}(t)}{\rho'}&=&\lb \frac{2(1+1/n)}{3\beta} \frac{V'cg}{\kappa R_*^2K\rho'^{1+1/n}} \rb^{1/(1+1/n+\beta)}
	t^{2/(1+1/n+\beta)}.
	\nonumber
	\\
	\label{eq:rhodiff_analytic}
\ee
For $n=3/2$, $\beta=0.19$, and using the equation of state for non-relativistic electrons, this gives
\be
	\rho_{0,\rm diff}(t) = 2.5\times10^6\kappa_{0.2}^{-0.54} E_{51}^{0.27}M_{1.4}^{0.37}R_{8.5}^{-2.46}t_{\rm day}^{1.08}\ {\rm g\ cm^{-3}},
	\nonumber
	\\
	\label{eq:rhodiff}
\ee
where $\kappa_{0.2}=\kappa/0.2\ {\rm cm^2\ g^{-1}}$, $E_{51}=E_{\rm sn}/10^{51}\ {\rm erg}$, $M_{1.4}=M_*/1.4M_\odot$, $R_{8.5}=R_*/3\times10^8\ {\rm cm}$, and $t_{\rm day} = t/1\ {\rm day}$. This density is roughly consistent with what one would expect for material dominated by non-relativistic electrons. To emphasize the depth in the WD this corresponds to, I estimate the diffusion mass as $M_{\rm diff}\approx (2/5)\times4\pi R_*^2\rho_{\rm 0,diff} H_0(\rho_{\rm 0,diff})$ (where the factor of 2/5 is appropriate for an $n=3/2$ polytrope), so that
\be
	M_{\rm diff}(t) \approx 1.5\times10^{-2}\kappa_{0.2}^{-0.90} E_{51}^{0.45}M_{1.4}^{-0.38}R_{8.5}^{-0.13}t_{\rm day}^{1.8}M_\odot.
	\label{eq:massdiff}
\ee
The most striking feature of these results is that  $^{56}$Ni must be present $\approx10^{-2}M_\odot$ below the surface at $\approx1\ {\rm day}$ post-explosion. The earliest detection of SN 2011fe is at $\approx11\ {\rm hr}$ (assuming a $t^2$ rise), which would require $^{56}$Ni at a depth of merely $\approx4\times10^{-3}M_\odot$.

\begin{figure}
\epsscale{1.2}
\plotone{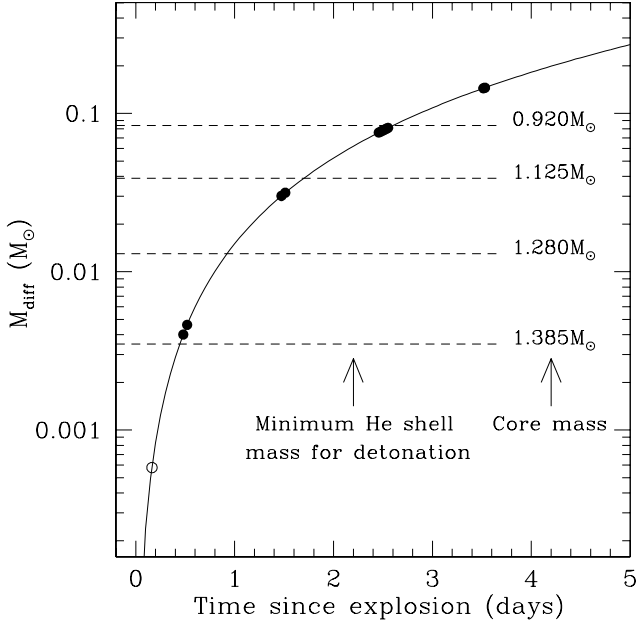}
\caption{The solid line marks the position of the thermal diffusion wave as a the surface layers expand using equation (\ref{eq:massdiff}). This shows the depth in the WD from which the luminosity is originating as a function of time. Along the solid line I have plotted filled circles at the time of the PTF observations \citep{nug11} and an open circle showing the time of an upper limit constraint \citep{blo12}. The horizontal dashed lines show the mimimum helium shell mass necessary for detonation, which are each labeled by their corresponding core mass \citep{fin10}. For example, for a $1.125M_\odot$ C/O core a $0.039M_\odot$ helium shell will successfully detonate, as will any larger mass helium shell.}
\label{fig:massdiff}
\epsscale{1.0}
\end{figure}
In Figure \ref{fig:massdiff}, I plot the depth of the diffusion wave below the WD surface as a function of time using equation (\ref{eq:massdiff}). The filled circles indicate the time of the PTF observations \citep{nug11}. Since the time of explosion of SN 2011fe is not exactly known, I assume it is 55796.687 MJD, which was derived assuming a $t^2$ rise \citep{blo12}. Also plotted as an open circle is the time of the upper limit reported by \citet{blo12}.

As a comparison, I also show the minimum helium shell masses needed for a surface detonation \citep[dashed lines, from][]{sb09,fin10}, which are labeled by their corresponding core masses (a smaller C/O core requires a larger helium shell mass for ignition). Such surface helium burning can ignite the core and create a double-detonation ignition. This comparison between the shell masses and the depth of the diffusion wave in SN 2011fe indicates  that a double-detonation ignition may provide radioactive ashes at the correct depth needed for powering the early-time rise. I further discuss the double-detonation scenario, along with other SNe Ia progenitor models, in \S \ref{sec:conclusions}.

\section{Rising Luminosity}
\label{sec:luminosity}

I next estimate the luminosity expected from a radioactively-dominated rising supernova, which will allow constraints to be placed on the amount and gradient of $^{56}$N near the WD surface. Combining equations (\ref{eq:radiativeluminosity}) and (\ref{eq:energydensity}), the luminosity from a given shell is
\be
	L(\rho_0,t) &=& \frac{1+1/n+3\beta+\chi}{\beta}\frac{4\pi cV'\epsilon_{56}X_{56}t^2}{3\kappa}
	\lp\frac{\rho_0}{\rho_{56}} \rp^\chi\lp \frac{\rho_0}{\rho'}\rp^{-\beta}.
	\nonumber
	\\
	\label{eq:luminosity}
\ee
From this result, a number of important features can already be identified. In the limit that $\chi$ and $\beta$ are small, all dependencies on $\rho_0$ are negligible, so that $L\sim cV't^2\epsilon_{56}/\kappa$, consistent with the single-zone result given by equation (\ref{eq:singlezone}).

To derive the luminosity as a function of time I substitute $\rho_{0,\rm diff}(t)$ from equation (\ref{eq:rhodiff_analytic}) in for $\rho_0$ in equation (\ref{eq:luminosity}),
\be
	L(t) &=& \frac{1+1/n+3\beta+\chi}{\beta}\lb \frac{2(1+1/n)}{3\beta}\frac{gV'c}{\kappa R_*^2K\rho'^{1+1/n}} \rb^{\frac{\chi-\beta}{1+1/n+\beta}}
	\nonumber
	\\
	&&\times
	\frac{4\pi cV'\epsilon_{56}X_{56}}{3\kappa}\lp\frac{\rho'}{\rho_{56}} \rp^{\chi}
	 t^{2(1+1/n+\chi)/(1+1/n+\beta)}.
	 \label{eq:luminosity2}
\ee
Again, when $\beta\approx\chi\approx 0$, the power law simplifies to $L\propto t^2$. As the gradient in $V_s$ becomes larger and $\beta$ increases, the power law decreases below 2. Conversely, a larger deposition index $\chi$ can increase the power law, since this means that more $^{56}$Ni is being probed as the diffusion wave moves deeper.

For the case of $n=3/2$, $\beta=0.19$, and $\chi=0$, the luminosity is
\be
	L(t) &=& 2.1\times10^{42}X_{56}\kappa_{0.2}^{-0.90} E_{51}^{0.45}M_{1.4}^{-0.38}R_{8.5}^{-0.10}t_{\rm day}^{1.8}\ {\rm erg\ s^{-1}}.
	\nonumber
	\\
	\label{eq:luminosity_t18}
\ee
For SN 2011fe, a luminosity of $\approx10^{40}\ {\rm erg\ s^{-1}}$ is seen at $\approx~11\ {\rm hr}$ \citep{nug11}. From the estimates given here, this requires a mass fraction of $X_{56}\sim2\times10^{-2}$ (although in the next section I make a more quantitative comparison to the observations). Such an amount of $^{56}$Ni is consistent with the energy density of the plasma being dominated by radioactive decay as discussed in Figure 1.

For my fiducial model given by equation (\ref{eq:luminosity_t18}), the rising luminosity scales like $t^{1.8}$ and not $t^2$. This is inconsistent with the power-law index of $2.01\pm0.01$ found by \citep{nug11}, but within the measurements of a broader range of SNe Ia \citep{con06}. This then begs the question, is an index of 2 universal across all SNe Ia, or can the power law vary around 2 depending on the gradients in velocity and radioactive heating for a given particular event? The work here argues via equation~(\ref{eq:luminosity2}) that a range of power-laws are possible. The fact that the best studied SNe Ia shows an index very close to 2 may be evidence that it is universal, but remember that the PTF data is in {\it g}-band. A detailed comparison with the {\it bolometric} rises of SNe Ia is necessary to understand how diverse the power-laws can actually be.

If the bolometric luminosity does indeed follow a simple $t^2$ scaling, then the $^{56}$Ni deposition must obey $\chi\approx\beta$. This cancels out the dependencies due to the position of the thermal diffusion wave (essentially the entire bracketed term in eq. [\ref{eq:luminosity2}] is set to 1), giving the simple dependencies that can be derived from a single-zone analysis,
\be
	L(t) &=& 2.7\times10^{42}X_{56,\rho_6}\kappa_{0.2}^{-1} E_{51}^{1/2}M_{1.4}^{-1/2}t_{\rm day}^{2}\ {\rm erg\ s^{-1}},
	\label{eq:luminosity_t2}
\ee
where now it is understood that $X_{56,\rho_6}$ is the mass fraction of $^{56}$Ni at $\rho_0=10^6\ {\rm g\ cm^{-3}}$, which corresponds to the depth of the diffusion wave at $\approx10\ {\rm hrs}$ (eq. [\ref{eq:rhodiff}]).

\section{Comparisons to SN 2011fe}
\label{sec:compare}

The effective temperature of the supernova during these early times may be complicated due to an opacity that depends on a mix of heavy elements. With this caveat in mind, I estimate the effective temperature assuming a constant, electron-scattering opacity. The effective temperature is
\be
	T_{\rm eff} = T/\tau^{1/4},
\ee
where $T= (U/a)^{1/4}$ is the local temperature and $\tau\approx\kappa \rho \Delta r$ is the optical depth, resulting in
\be
	T_{\rm eff} = \lp \frac{1+1/n}{4\beta} \frac{\epsilon_{56}X_{56}}{a\kappa V'}\rp^{1/4}
	\lp\frac{\rho_0}{\rho_{56}} \rp^{\chi/4}
	\lp \frac{\rho_0}{\rho'}\rp^{\beta/4}.
\ee
Evaluating this at the diffusion depth provides an estimate for the time evolution of the effective temperature. For $n=3/2$, $\beta=0.19$, and $\chi=0$,
\be
	T_{\rm eff}(t) = 1.6\times10^4X_{56}^{1/4}\kappa_{0.2}^{-0.28} E_{51}^{-0.11}M_{1.4}^{0.10}R_{8.5}^{0.03}t_{\rm day}^{0.05} \ {\rm K}.
	\label{eq:teff1}
\ee
This shows that the radioactive heating balances the expansion to give an effective temperature that is nearly constant with time.

\begin{figure}
\epsscale{1.2}
\plotone{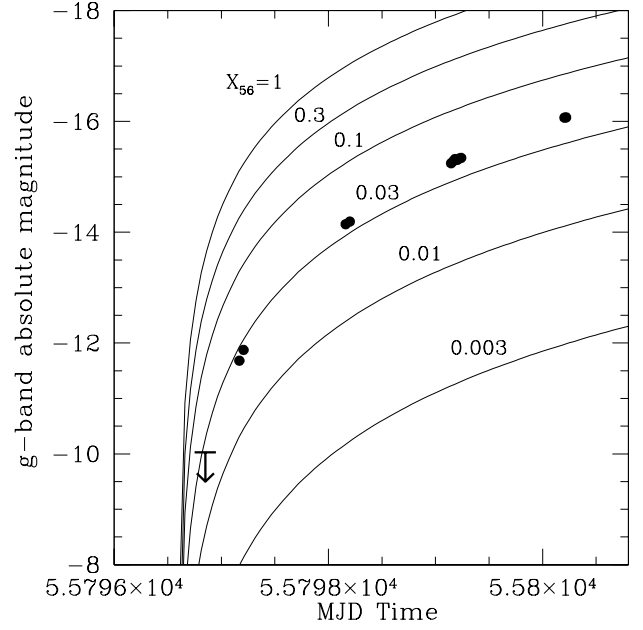}
\caption{Rising lightcurve for a model with $\beta=0.19$, $n=3/2$, $\chi=0$, $E_{\rm sn}=10^{51}\ {\rm erg}$, $M_*=1.4M_\odot$, and $R_*=3\times10^8\ {\rm cm}$. The $^{56}$Ni is varied from $X_{56}=3\times10^{-3}$ to $1$, as labeled on each curve. Although $X_{56}\approx 3\times10^{-2}$ is favored, this model with a constant distribution of $^{56}$Ni overpredicts the early lightcurve and underpredicts the late lightcurve. This indicates that a gradient in $^{56}$Ni deposition is needed.}
\label{fig:lightcurve}
\epsscale{1.0}
\end{figure}

The photospheric radius $r_{\rm ph}$ of the expanding ejecta can be estimated with relationship $L=4\pi r_{\rm ph}^2 \sigma_{\rm SB}T_{\rm eff}^4$, where $\sigma_{\rm SB}$ is the Stefan-Boltzmann constant. When the diffusion wave is at $\rho_0$, the photospheric radius is therefore
\be
	r_{\rm ph}(\rho_0,t) = \lb \frac{1+1/n+3\beta+\chi}{3(1+1/n)}\rb^{1/2} 4V' t\lp\frac{\rho_0}{\rho'} \rp^{-\beta},
\ee
which for $n=3/2$, $\beta=0.19$, and $\chi=0$, and using equation~(\ref{eq:rhodiff}), results in
\be
	r_{\rm ph}(t) = 2.2\times10^{14}\kappa_{0.2}^{0.10} E_{51}^{0.45}M_{1.4}^{-0.38}R_{8.5}^{0.10}t_{\rm day}^{0.80} \ {\rm cm}.
	\label{eq:rph}
\ee	
Using this photospheric radius, and assuming blackbody emission with effective temperature given by equation (\ref{eq:teff1}), I calculate the {\it g}-band absolute magnitude in Figure \ref{fig:lightcurve}. Plotted as solid circles are the observations summarized by \citet{nug11}, along with an arrow indicating the upper limit from \citet{blo12}. For this comparison a choice of explosion time is necessary, for which I use 55796.6 MJD. This is not meant to maximize the fit between theory and data, but merely to provide a useful comparison. Although $X_{56}\approx 3\times10^{-2}$ is favored, this model overpredicts the early lightcurve and underpredicts the late lightcurve, indicating that a gradient in $^{56}$Ni deposition is needed to make a better match. This is not surprising since these models correspond to $L\propto t^{1.8}$ and not $t^2$.

\begin{figure}
\epsscale{1.2}
\plotone{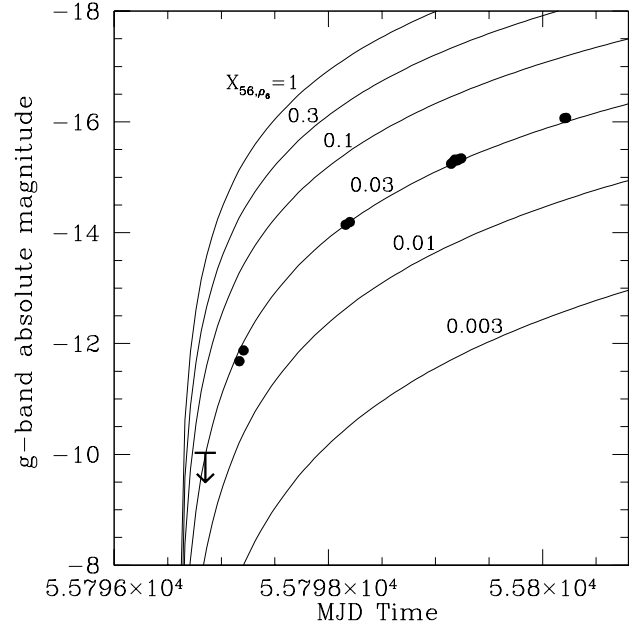}
\caption{The same as Figure \ref{fig:lightcurve}, but with a $^{56}$Ni gradient set to $\chi=\beta=0.19$ instead. The model with $X_{56,\rho_6}\approx 3\times10^{-2}$ provides a reasonable fit to the observed lightcurve.}
\label{fig:lightcurve2}
\epsscale{1.0}
\end{figure}
Motivated by this discrepancy between the observed lightcurve and that predicted for a constant distribution of $^{56}$Ni, I instead consider the case $\chi=\beta=0.19$, which was found to give a $t^2$ bolometric luminosity (eq. [\ref{eq:luminosity_t2}]). The corresponding effective temperature is
\be
	T_{\rm eff}(t) = 1.6\times10^4X_{56,\rho_6}^{1/4}\kappa_{0.2}^{-0.30} E_{51}^{-0.10}M_{1.4}^{0.08}R_{8.5}^{-0.09}t_{\rm day}^{0.10} \ {\rm K},
	\label{eq:teff2}
\ee
where again I emphasize that $X_{56,\rho_6}$ is the mass fraction of $^{56}$Ni at a depth of $\rho_0=10^6\ {\rm g\ cm^{-3}}$. The difference between equations (\ref{eq:teff1}) and (\ref{eq:teff2}) may not seem appreciable, but remember that $L\propto T_{\rm eff}^4$, so that small changes in exponents can make an important difference. The photospheric radius is essentially unchanged from equation (\ref{eq:rph}), with only a small difference in the prefactor.

The resulting lightcurves for the $\chi=0.19$ model is plotted in Figure \ref{fig:lightcurve2}, which demonstrates a better fit to the data. The necessary gradient in $^{56}$Ni implies that $X_{56}\approx 3\times10^{-2}$ at a depth of $M_{\rm diff}\approx 4\times10^{-3}M_\odot$ and this varies up to $X_{56}\approx 5\times10^{-2}$ at a depth of $M_{\rm diff}\approx 0.2M_\odot$. The total integrated mass of radioactive material needed to produce the lightcurve over $\approx4\ {\rm days}$ is roughly $\approx10^{-2}M_\odot$.

\section{Conclusions and Discussion}
\label{sec:conclusions}

I investigated the impact of radioactive heating on the early-time rise of SN Ia lightcurves. A mass fraction $X_{56}\approx3\times10^{-2}$ of $^{56}$Ni at a depth $\approx4\times10^{-3}-0.1M_\odot$ from the progenitor surface is needed to produce the luminosity seen from SN 2011fe during its first $\approx4\ {\rm days}$ (although, as discussed in \S 2, other heating sources such as $^{48}$Cr could just as well explain the rise). A model in which there is a velocity gradient set by the passage of a shock, but with a constant deposition of radioactive material, gives a $t^{1.8}$ power law. This appears to be inconsistent with SN 2011fe, but is not ruled out by the stacking of many nearby SNe Ia \citep{con06}. The shape of the SN 2011fe lightcurve is better fit when the gradients in velocity and radioactive material are similar, with $\chi\approx\beta$, as is shown in Figure \ref{fig:lightcurve2}. In the future, a comparison between the {\it bolometric} rising luminosity of SN 2011fe and this work using equation (\ref{eq:luminosity2}) would provide important constraints on the distribution of radioactive material near the WD surface.

A crucial question for future observations is whether the rise of SNe Ia obey a universal power-law or if they vary from event to event. My work shows that the rise should depend on the particular gradients in velocity, density, and deposition of radioactive elements, which, although it may be close to $t^2$, should not necessarily always be the same. If the power law is indeed found to be universal, this would argue that different physics than what I am exploring here is determining the rise. For example, an opacity effect that I have not included could enforce a fixed $T_{\rm eff}$ and constant expansion velocity.

The total integrated mass of radioactive material needed to produce the SN 2011fe lightcurve over $\approx4\ {\rm days}$ is $\approx10^{-2}M_\odot$. Single detonation models of Chandrasekhar and even sub-Chandrasekhar-mass WDs find distributions of ashes that are fairly well-stratified, and do not show $^{56}$Ni or $^{48}$Cr near the surface. This is perhaps not that limiting of a constraint; Chandrasekhar-mass single detonations are disfavored because they cannot produce the observed intermediate mass elements \citep{fil97}, and although sub-Chandrasekhar single detonations match the nucleosynthesis generally seen from SNe Ia \citep{sim10}, it is not known how to ignite such an object without a helium shell.

Conversely, the explosive ignition of a helium shell in the double-detonation scenario can produce shallow radioactive material. Indeed, Figure 2 shows that the depth of this material is roughly the same as the minimum helium shell masses needed for detonation \citep{sb09,fin10}. The total amount of radioactive material needed is also qualitatively similar to the nucleosynthesis of such events \citep{fin10}. DDT models can produce $^{56}$Ni near the WD surface, but potentially only in the most strongly mixed, off-center deflagrations \citep{mae10}. In DDT models with many ignition points that have fairly stratified ashes, radioactive elements are not present near the surface. A gravitationally confined detonation (GCD) is in a sense just a more extreme, off-center version of the DDT models, and it too produces iron-peak elements near the surface when a bubble unstably rises and breaks at the top \citep{mea09}. Finally, a more speculative idea is that some radioactive elements are synthesized near the surface by {\it g}-mode heating during the pre-explosive convective phase \citep{pir11}. \citet{nug11} report the presence of O, Mg, Si, S, Ca, and Fe in the spectra of SN 2011fe at early times. Many of these elements are potential ashes from the scenarios described above, and a more detailed comparison may help discriminate between them.

Even though this work argues for the presence of radioactive material near the surface of the WD, a potential problem is that if the abundance of iron-peak elements is too high, they tend to produce colors that are too red and spectra that are inconsistent with normal Type Ia supernovae \citep{kro10,sim11}. Although it should be noted that this difficulty may be partially alleviated for larger mass WDs that have smaller helium shells, and may also depend on careful consideration of the burning in the helium shell \citep{wk11}. Combining detailed modeling of the early lightcurve rise with comparisons to the peak colors and spectra provides competing limits on the mass fraction of iron-peak elements, and therefore together should result in tight constraints on the composition of the outer layers.
\\

I thank Lars Bildsten, Luc Dessart, Dan Kasen, Peter Goldreich, Christian Ott, and Ken Shen for helpful comments and discussions. This work was supported through NSF grant AST-0855535 and by the Sherman Fairchild Foundation.

\begin{appendix}

\section{General Self-Similar Rising Luminosity Solutions}

Here I consider a wider range of self-similar solutions for the rising luminosity. This highlights the expected changes in the time-dependence as various details are added to future models.

\subsection{General Opacity Law}

For the majority of this work I have assumed a constant opacity, consistent with electron scattering. To make better comparisons with observations will require complete calculation with detailed opacities. If helium is present near the WD surface (such as in the double-detonation scenario), it may recombine in expanded, cooled layers. Metals lines would provide a strong opacity in the UV. In light of this, I consider a more general opacity law
\be
	\kappa = \kappa_0 \rho^a T^b.
\ee
The self-similar solution for such a case results in $L\propto t^\lambda$ (analogous to eq. [\ref{eq:luminosity2}]), where
\be
	\lambda = \lp 2+3a+b/2 \rp\lb \frac{1+1/n+\chi +\chi b /2}{1+1/n+\beta +(1+1/n+3\beta)(a+b/4)+\chi b /4}  \rb.
\ee
For example, for a pure C/O composition \citet{rw11} use $\kappa=0.66(T/10^5\ {\rm K})^{1.27}\ {\rm cm^2\ g^{-1}}$. Combining such an opacity law with $n=3/2$, $\beta=0.19$, and $\chi=0$ results in $\lambda = 1.71$ for the power-law exponent. When instead $\chi=\beta=0.19$, then $\lambda=1.98$, which shows that a roughly quadratic luminosity increase is fairly robust as long as the nickel deposition is increasing.


\subsection{Non-Plane-Parallel Solution}

In the continuity relation, equation (\ref{eq:massconservation}), I made the plane-parallel assumption that essentially all the exploding material came from roughly the same radius of $R_*$. This is a reasonable approximation for a compact progenitor like a WD. Conversely, for a more extended progenitor continuity becomes
\be
	\rho(\rho_0,t) = \rho_0 \lb\frac{r_0}{V(\rho_0)t} \rb^2 \lp \frac{H_0}{\Delta V(\rho_0) t} \rp,
\ee
where $r_0$ is the initial radius for a given shell of material. For a polytrope, this scales $r_0\propto \rho_0^{-1/n}$ at depths well below the surface of the star. Completing the self-similar analysis with a constant opacity, the diffusion wave now has a time-dependent position of
\be
	\rho_{0,\rm diff}(t) \propto t^{2/(1-1/n+\beta)}.
\ee
This generally gives a larger exponent than the plane-parallel case in equation (\ref{eq:rhodiff_analytic}), showing that the diffusion wave traverses through the star more rapidly in this case. The time dependence of the luminosity is
\be
	L \propto t^{2(1-1/n+\chi)/(1-1/n+\beta)}.
\ee
For $n=3/2$, $\beta=0.19$ and $\chi=0$, this results in an exponent of 1.27, which is much below 2. This means that as non-plane-parallel effects become more important, the luminosity will begin to flatten. Also note that this new exponent is much more sensitive to $\chi$, since setting $\chi=\beta$ increases the exponent all the way up to 2. These non-plane-parallel solution may also have use in investigating Type Ib/c supernovae at early enough times that the lightcurve is still rising, but not so early that the emission if dominated by the shock-heating of the surface layers \citep[as studied by][]{ns10}.

\end{appendix}

\end{document}